\documentclass[prd,amsmath,amssymb,preprintnumbers,nofootinbib,reprint,notitlepage]{revtex4-1}
\pdfoutput=1
\usepackage[utf8]{inputenc} 
\usepackage{graphicx}
\usepackage{url}
\usepackage[bookmarks, pagebackref=false]{hyperref}
\usepackage[usenames,dvipsnames]{xcolor}
\usepackage{cleveref}

\definecolor{orange}{cmyk}{0,0.5,1,0}
\definecolor{graa}{rgb}{0.8,0.8,0.8}
\definecolor{blaa}{rgb}{0.2,0.2,0.6}

\definecolor{colA}{HTML}{c19277}
\definecolor{colB}{HTML}{e1bc91}
\definecolor{colD}{HTML}{62959c}

\hypersetup{
  colorlinks, 
  bookmarksopen, 
  bookmarksnumbered,
  citecolor=colA, 		
  linkcolor=colA,	
  urlcolor=colD,			
}
\usepackage{amsthm}
\usepackage{bm}
\usepackage{bbm}
\usepackage{pxfonts}

\usepackage{amsmath,amssymb,amsfonts}
\usepackage{color}
\usepackage{hyperref}
\usepackage[Symbolsmallscale]{upgreek}
\usepackage{amsmath}
\usepackage{amsfonts}
\usepackage{amssymb,dsfont}
\usepackage{graphicx}
\usepackage{amssymb}

\usepackage{placeins}
\usepackage{xspace}
\usepackage{cancel} 

\usepackage{slashed}

\usepackage{natbib}

\newcommand{\ep}{\ensuremath{\varepsilon}}
\newcommand{\Zmu}{\ensuremath{Z_{\mu}}}
\newcommand{\Zlam}{\ensuremath{Z_{\lambda}}}
\newcommand{\Dmu}{\ensuremath{\partial_{\mu}}}
\newcommand{\Dlam}{\ensuremath{\partial_{\lambda}}}

\newcommand{\msmu}{\ensuremath{\mathfrak{m}}}
\newcommand{\MS}{\ensuremath{\mathrm{\overline{MS}}}}
\newcommand{\hvec}{\ensuremath{\mathbf{h}}}

\newcommand{\denA}{\ensuremath{\textcolor{colD}{\left[ 20+d_c \right]}}}
\newcommand{\denB}{\ensuremath{\textcolor{colD}{\left[ 24+d_c \right]}}}

\begin{document}

\title{
  \Large\color{colD} 
  Four-loop critical properties of polymerized membranes}

\author{\sc Andrey Pikelner}\email{pikelner@theor.jinr.ru} 
\affiliation{Bogoliubov Laboratory of Theoretical Physics, Joint Institute for
  Nuclear Research, 141980~Dubna, Russia}

\begin{abstract}
  We calculate four-loop order corrections to the critical exponent $\eta$ in
  the two-field model of flat phase membranes. Obtained results show better
  agreement with the other calculation methods and confirm the validity of the
  perturbative approach to the considered problem.
\end{abstract}

\maketitle

\section{Introduction}
\label{sec:intro}
Description of many real-life systems can be reduced to the models of
polymerized $D$-dimensional membranes in their low-temperature flat phase embedded
in the $d$-dimensional Euclidean space. Using renormalization group methods, one can
describe the flow of the theory between non-interacting UV-stable Gaussian fixed
point and IR-attractive flat phase fixed point. Many different approaches
describe the model near the IR fixed-point, non-perturbative such as
NPRG~\cite{PhysRevE.79.040101}, self-consistent screening
approximation (SCSA)~\cite{LeDoussal:1992xh,PhysRevE.80.041117,LEDOUSSAL2018340}, numerical SCSA~\cite{PhysRevB.82.125435,PhysRevB.83.174104} and
Monte Carlo simulations~\cite{PhysRevE.48.R651,Bowick:1996wz,PhysRevB.87.104112}. Also, perturbative methods based on small epsilon expansion have been successfully applied to the problem starting from pioneering works~\cite{NelsonPeliti,Aronovitz:1988zz} and recently extended to three-loop order in the
series of papers~\cite{Coquand:2020tgb,Mauri:2021ili,Metayer:2021kxm}. The main object of interest is to estimate critical exponents
controlling power-law behavior of the phonon-phonon and flexural-flexural
correlation functions near the critical point.

The recent three-loop calculation~\cite{Metayer:2021kxm} provides us with a new value of critical
exponent very close to the results from other methods. Moreover, it demonstrates
an apparent convergence of the perturbative series in the proper direction. The present paper
aims to perform four-loop calculations and check perturbative series behavior
for the critical exponent at the next loop order.

We carry out all our calculations in the two-field model described in detail in~\cite{Aronovitz:1988zz}
with action given by:
\begin{align}
  \label{eq:S-2field-def}
  S =\frac{1}{2}\int d^D x\left[(\partial^2 \hvec)^2 + 2 \mu\, u_{ij}^2 + \lambda\,u_{ii}^2\right],\\
  u_{ij} = \frac{1}{2}\left( \partial_iu_j +\partial_j u_i + \partial_i \hvec \cdot \partial_j \hvec\right)
\end{align}
Here we keep only relevant part of $u_{ij}$ and appropriately
rescaled\footnote{To simplify notations, we use the model in such a form that fields
  and Lame coefficients are normalized by $\kappa$ and use simply $\lambda$ and $\mu$ for
  quantities divided by $\kappa$.} fields $u,\hvec$ and Lame coefficients
$\lambda,\mu$. Our goal is to calculate renormalization constants of fields and
couplings entering~\eqref{eq:S-2field-def} as expansion about the upper
critical dimension $D_{\rm uc} = 4$ in a small parameter $\ep = 2-D/2$. Zeroes of
beta-functions derived from the renormalization constant provide us with the
coordinates of the set of fixed points $\{\mu^*, \lambda^*\}$ which after
substitution into the field anomalous dimension lead to the final answer for the
critical exponent of the interest.

Our main result is the a new value of $\eta$ for the $d_c=1$ case corresponding to
the $D=2$ membrane embedded into the $d=3$ dimensional space at the four-loop order:
\begin{equation}
  \label{eq:etaP4res}
  \eta = 0.8670
\end{equation}

In section~\ref{sec:calc-details}, we provide details of our four-loop
calculation and in section~\ref{sec:results} we make a comparison with
available results and present analytical expressions for obtained results
together with cross checks.
\section{Calculation details}
\label{sec:calc-details}
The possibility of highly nontrivial four-loop calculations described below
is based on two essential features of the considered model:
\begin{itemize}
\item In all our calculations, we are allowed to consider massless diagrams
  only.
\item Thanks to the Ward identities, all needed renormalization constants can
  be derived from two-point functions only.
\end{itemize}
Recent two-loop~\cite{Coquand:2020tgb} and three-loop~\cite{Metayer:2021kxm}
calculations successfully adopted both of these facts. Table~\ref{tab:n-dias}
provides us with the number of diagrams that need to be calculated up to the
four-loop level, which is an order of magnitude higher than in the three-loop
problem considered before~\cite{Metayer:2021kxm}. Thus, our main improvement
consists of applying modern tools of multi-loop calculations and enhanced
renormalization strategy compared to~\cite{Coquand:2020tgb,Metayer:2021kxm}.
\begin{table}[h]
  \centering
  \setlength{\tabcolsep}{10pt}
  \renewcommand*{\arraystretch}{1.2}  
  \begin{tabular}{rcccc}
    & 1-loop & 2-loop & 3-loop & 4-loop\\
    $\Pi_{hh}$ & 1 & 6 & 45 & 516 \\
    $\Pi_{uu}$         & 1 & 3 & 23 & 237
  \end{tabular}
  \caption{Number of calculated diagrams}
  \label{tab:n-dias}
\end{table}
Feynman rules for the model~\eqref{eq:S-2field-def} used in our calculation can
be found in~\cite{guitter:1989tbpm}. We generate all diagrams with
\texttt{DIANA}~\cite{Tentyukov:1999is} and calculate all massless integrals up
to the four loops with \texttt{FORCER}~\cite{Ruijl:2017cxj}. Applying
appropriate projectors we obtain bare results for the sum of one-particle
irreducible(1PI) diagrams. In case of the $\hvec$-field the projector is trivial
\begin{equation}
  \label{eq:Piuu-proj}
  \Pi_{a,b}^{hh} = \delta_{a,b} \Pi_{hh}, \quad a,b=1\dots d_c
\end{equation}
with $d_c = d-D$ but in the case of the $u$-field we split result of the
calculation into transverse and longitudial parts:
\begin{equation}
  \label{eq:Piuu-proj}
  \Pi_{i,j}^{uu} = \left(\delta_{i,j} - \frac{Q_i\cdot Q_j}{Q^2} \right)\Pi_{uu}^T + \frac{Q_i\cdot Q_j}{Q^2}\Pi_{uu}^L,
\end{equation}
where $i,j = 1\dots D$. Bare quantities\footnote{We utilize subscript B for bare
quantities.} are related to renormalized ones by $h_{B} = \sqrt{Z} h$, $u_B = Z
u$, $\lambda_B = \Zlam\lambda$, $\mu_B = \Zmu\mu$.
Renormalizability of the theory allows us to greatly simplify renormalization
of the theory and extraction of constants $Z,\Zmu,\Zlam$.
Instead of explicit account of diagrams with counterterms insertions or any
other ways of subtraction of UV divergencies like BPHZ, it is possible to renormalize
couplings in the bare result for the calculated Green functions and multiply it
with an appropriate renormalization constant. Similar approach was pioneered in the
three-loop QCD renormalization~\cite{Larin:1993tp} and proved to be especially
useful for renormalization of the Standard Model~\cite{Bednyakov:2012rb}.
Our starting point is the calculation of the sum of bare 1PI diagrams up to the
four-loop order denoted as $\Pi_{hh}$ for the $\hvec$-field and
$\Pi_{uu}^T,\Pi_{uu}^L$ for transverse and longitudinal parts in the case of the
$u$-field respectively. Here it is important to keep terms in $\ep$-expansion up
to the order $\ep^{3-L}$ in the calculation of the $L$-loop diagrams. The sum of
1PI diagrams with an appropriate tree term after replacement of bare couplings
with renormalized ones and multiplication by the overall renormalization
constant is finite and we have three equations to fix three renormalization
constants $Z,\Zmu,\Zlam$:
\begin{align}
  & Z^2 \left(\mu \Zmu - \Pi_{uu}^T(\mu_B \to \mu \Zmu, \lambda_B \to \lambda \Zlam)\right) = \mathcal{O}(\ep^0) 
    \nonumber\\
  & Z^2 \left(2 \mu \Zmu + \lambda \Zlam - \Pi_{uu}^L(\mu_B \to \mu \Zmu, \lambda_B \to \lambda \Zlam)\right) = \mathcal{O}(\ep^0)
    \nonumber\\
  & Z \left(1 - \Pi_{hh}(\mu_B \to \mu \Zmu, \lambda_B \to \lambda \Zlam)\right) = \mathcal{O}(\ep^0)   \label{eq:Pi-ren}
\end{align}
Due to the complicated dependence of $Z_i$ on couplings $\mu,\lambda$, it is
usefull to introduce the loop counting parameter and solve equations
\eqref{eq:Pi-ren} perturbatively order by order to get four-loop renormalization
constants $Z,\Zmu,\Zlam$\footnote{All results are available in a
computer-readable form as ancillary files to the arxiv version of the paper.}.
Beta functions $\beta_X=\frac{\partial X}{\partial \log \msmu}$ and field
anomalous dimension $\gamma=\frac{\partial \log Z}{\partial \log \msmu}$ are
defined as logarithmic derivatives in $\MS$ scale parameter $\msmu$. From
calculated $Z,\Zmu,\Zlam$ we can find beta functions:
\begin{align}
  \beta_{\mu} & = \frac{2\ep \partial_{\lambda}\log{\frac{\mu \Zmu}{\lambda\Zlam}}}
  {\det \begin{pmatrix}
      \Dlam \log{\lambda \Zlam} & \Dmu \log{\lambda \Zlam} \\
      \Dlam \log{\mu \Zmu} & \Dmu \log{\mu \Zmu}
    \end{pmatrix}
                             } \label{eq:betaMU-compact}\\
  \beta_{\lambda} &= \frac{2\ep \partial_{\mu}\log{\frac{\lambda\Zlam}{\mu \Zmu}}}
  {\det \begin{pmatrix}
      \Dlam \log{\lambda \Zlam} & \Dmu \log{\lambda \Zlam} \\
      \Dlam \log{\mu \Zmu} & \Dmu \log{\mu \Zmu}
    \end{pmatrix}
                             }  \label{eq:betaLAM-compact}
\end{align}
and the field anomalous dimension
\begin{equation}
  \label{eq:gamPhi-def}
  \gamma = \beta_{\lambda} \Dlam \log{Z} + \beta_{\mu} \Dmu \log{Z}
\end{equation}
The absense of $\ep$ poles in
\eqref{eq:betaMU-compact},\eqref{eq:betaLAM-compact} and \eqref{eq:gamPhi-def}
is a strong indication of the validity of the obtained results.

\section{Results}
\label{sec:results}
From the set of equations $\beta_{\mu}(\mu^*, \lambda^*) =0,
\beta_{\lambda}(\mu^*, \lambda^*)=0$ we have found four different solutions
$(\mu^*, \lambda^*)$ corresponding to four different fixed points. We adopt the
same notation as in~\cite{Aronovitz:1988zz}, where the point $P_1$ is the
Gaussian one, and $P_4$ is IR attractive one we are interested in. In addition,
we consider unstable fixed point $P_3$, since it allows comparing the result of
the calculations with $1/d_c$ results available in the literature. To fix the
notation and simplify comparison with \cite{Coquand:2020tgb,Metayer:2021kxm}, we
provide one-loop coordinates of the points $P_3$ and $P_4$:
\begin{alignat}{2}
  & \mu_3^* = \frac{12}{20 + d_c}\ep + \mathcal{O}(\ep^2) \qquad
  && \lambda_3^* = -\frac{6}{20 + d_c}\ep + \mathcal{O}(\ep^2) \\
  & \mu_4^* = \frac{12}{24 + d_c}\ep + \mathcal{O}(\ep^2) \qquad
  && \lambda_4^* = - \frac{4}{24 + d_c}\ep + \mathcal{O}(\ep^2)
\end{alignat}
Substituting four-loop results for fixed-point coordinates into four-loop field
anomalous dimension~\eqref{eq:gamPhi-def}, we obtain critical exponents $\eta =
\gamma(\mu^*, \lambda^*)$ for two selected fixed points.
Full analytic four-loop results for the critical exponents $\eta_3$ (fixed point
$P_3$) and $\eta_4$ (fixed point $P_4$) can be found in the Appendix in
Eq.\eqref{eq:eta3-4loop} and Eq.\eqref{eq:eta4-4loop} respectively. For the case
$d_c=1$ our result reads:
\begin{align}
  \eta_3 = & 0.952 \ep - 0.071 \ep^2 - 0.069 \ep^3 
           - 0.075 \ep^4 + \mathcal{O}(\ep^5)  \label{eq:eta3Numdc1}\\
  \eta_4 = & 0.96 \ep - 0.0461 \ep^2 - 0.0267 \ep^3
           - 0.02 \ep^4 + \mathcal{O}(\ep^5)\label{eq:eta4Numdc1}
\end{align}
The three-loop parts of \eqref{eq:eta3Numdc1} and \eqref{eq:eta4Numdc1} coincide
with result of three-loop calculation\cite{Metayer:2021kxm} and the four-loop
term is new. For the important case $D=2$, substituting $\ep=1$ into
\eqref{eq:eta3Numdc1} and \eqref{eq:eta4Numdc1} we obtain our final result for
the critical exponent $\eta$. Comparison with the earlier
two-loop~\cite{Coquand:2020tgb} and three-loop\cite{Metayer:2021kxm}
calculations and also with nonperturbative results obtained with NPRG
technique~\cite{PhysRevE.79.040101}, SCSA~\cite{LeDoussal:1992xh} and from MC
simulation~\cite{PhysRevB.87.104112} are summarized in table~\ref{tab:nu-loop}.
\begin{table*}
  \centering
  \setlength{\tabcolsep}{10pt}
  \renewcommand*{\arraystretch}{1.2}  
  \begin{tabular}{ccccccccc}
    & 1-loop & 2-loop & 3-loop & 4-loop & $[2/2]$ & SCSA~\cite{LeDoussal:1992xh} & NPRG~\cite{PhysRevE.79.040101} & MC~\cite{PhysRevB.87.104112}\\\hline
    $\eta_3$    &  0.9524 & 0.8813 & 0.8116 & 0.7368 &  -   &    -    &  -   & - \\\hline
    $\eta_4$    &  0.96     & 0.9139 & 0.8872 & 0.8670 & 0.806 & 0.821  & 0.849  & 0.795(10)
  \end{tabular}
  \caption{Perturbative results for $\eta$ compared to other methods predictions
  for the $D=2$ case}
  \label{tab:nu-loop}
\end{table*}
The obtained result provides additional support for apparent convergence of the
series for $\eta$ near $D=2$ without any additional resummation. As the trivial
exercise, we also construct a $[2/2]$ Pade aproximant for the 4-loop result
$\eta_4^{[2/2]} = 0.806$, which provides us with a slightly different result,
and indicates on the possible need for more careful resummation of the series.
The growth of the last expansion term in~\eqref{eq:eta3Numdc1} compared to the
three-loop one implies the possibility of the divergence of the series and
motivates for further more careful resummation of obtained series.
Another check of the validity of the obtained perturbative series stems from the
comparison with known $1/d_c$ expansions in the vicinity of chosen critical
points. Our main interest is the expression for $\eta_4$ but the result for
$\eta_3$ is also important, since it's leading order expansion in $1/d_c$ also
can be verified with results available in the literature.
In \cite{LeDoussal:1992xh}, the following result for the leading term in $1/d_c$
expansion corresponding to the fixed point $P_4$ was derived:
\begin{equation}
  \label{eq:eta-large-dc}
  \eta(D,d_c) = \frac{8}{d_c}\frac{D-1}{D+2}\frac{\Gamma(D)}{\Gamma(D/2)^3\Gamma(2-D/2)} + \mathcal{O}\left( \frac{1}{d_c^2} \right)
\end{equation}
after expansion in $\ep = 2-D/2$ up to the $\ep^4$, it perfectly matches the
leading term of \eqref{eq:eta4-4loop} expansion in $1/d_c$ and reads:
\begin{align}
  \label{eq:eta4-large-d}
  \eta_4 =
           \frac{1}{d_c}& \left(24 \ep - 24 \ep^2 - \frac{64}{3} \ep^3\right.\nonumber\\
  & \left.- \frac{16}{9} (13 - 27 \zeta_3)\ep^4 + \mathcal{O}(\ep^5) \right)
  + \mathcal{O}\left( \frac{1}{d_c^2} \right)
\end{align}
For the fixed point $P_3$ according to~\cite{LEDOUSSAL2018340} leading term of
the $1/d_c$ expansion is determined by $\eta\left(D, \frac{D (D - 1)}{(D - 2)(D
+ 1)} d_c\right)$, and after expansion in $\ep$ becomes equal to the $1/d_c$
expansion of \eqref{eq:eta3-4loop}:
\begin{align}
  \label{eq:eta3-large-d}
  \eta_3 =
  \frac{1}{d_c}& \left(  20 \ep - \frac{74}{3} \ep^2 - \frac{155}{9} \ep^3 \right.\nonumber\\
               & \left.-  \frac{1}{54}(769 - 2160 \zeta_3) \ep^4 + \mathcal{O}(\ep^5)  \right)
  + \mathcal{O}\left( \frac{1}{d_c^2} \right)
\end{align}

\section{Conclusion}
\label{sec:concl}
We have calculated four-loop beta-functions and field anomalous dimension in the
two-field model of polymerized membranes. From the obtained perturbative
results, we have found a set of fixed points and derived the value of the
critical exponent $\eta$ for $D=2$ case. The obtained results are in good
agreement with other calculation methods and demonstrate apparent convergence of
the perturbative series. The validity of the result is confirmed by comparison
with known results in $1/d_c$ expansion.

\acknowledgments
We thank A.Bednyakov, N.Lebedev, G.Kalagov, and M.Kompaniets for fruitful
discussions and useful comments on the manuscript. Comments and discussions
regarding the work~\cite{Metayer:2021kxm} by S.Metayer and S.Teber are kindly
acknowledged. Furthermore, we are grateful to the Joint Institute for Nuclear
Research for using their supercomputer ``Govorun.''
\bibliography{membranes4l}

\begin{thebibliography}{19}%
\makeatletter
\providecommand \@ifxundefined [1]{%
 \@ifx{#1\undefined}
}%
\providecommand \@ifnum [1]{%
 \ifnum #1\expandafter \@firstoftwo
 \else \expandafter \@secondoftwo
 \fi
}%
\providecommand \@ifx [1]{%
 \ifx #1\expandafter \@firstoftwo
 \else \expandafter \@secondoftwo
 \fi
}%
\providecommand \natexlab [1]{#1}%
\providecommand \enquote  [1]{``#1''}%
\providecommand \bibnamefont  [1]{#1}%
\providecommand \bibfnamefont [1]{#1}%
\providecommand \citenamefont [1]{#1}%
\providecommand \href@noop [0]{\@secondoftwo}%
\providecommand \href [0]{\begingroup \@sanitize@url \@href}%
\providecommand \@href[1]{\@@startlink{#1}\@@href}%
\providecommand \@@href[1]{\endgroup#1\@@endlink}%
\providecommand \@sanitize@url [0]{\catcode `\\12\catcode `\$12\catcode
  `\&12\catcode `\#12\catcode `\^12\catcode `\_12\catcode `\%12\relax}%
\providecommand \@@startlink[1]{}%
\providecommand \@@endlink[0]{}%
\providecommand \url  [0]{\begingroup\@sanitize@url \@url }%
\providecommand \@url [1]{\endgroup\@href {#1}{\urlprefix }}%
\providecommand \urlprefix  [0]{URL }%
\providecommand \Eprint [0]{\href }%
\providecommand \doibase [0]{http://dx.doi.org/}%
\providecommand \selectlanguage [0]{\@gobble}%
\providecommand \bibinfo  [0]{\@secondoftwo}%
\providecommand \bibfield  [0]{\@secondoftwo}%
\providecommand \translation [1]{[#1]}%
\providecommand \BibitemOpen [0]{}%
\providecommand \bibitemStop [0]{}%
\providecommand \bibitemNoStop [0]{.\EOS\space}%
\providecommand \EOS [0]{\spacefactor3000\relax}%
\providecommand \BibitemShut  [1]{\csname bibitem#1\endcsname}%
\let\auto@bib@innerbib\@empty
\bibitem [{\citenamefont {Kownacki}\ and\ \citenamefont
  {Mouhanna}(2009)}]{PhysRevE.79.040101}%
  \BibitemOpen
  \bibfield  {author} {\bibinfo {author} {\bibfnamefont {J.-P.}\ \bibnamefont
  {Kownacki}}\ and\ \bibinfo {author} {\bibfnamefont {D.}~\bibnamefont
  {Mouhanna}},\ }\href {\doibase 10.1103/PhysRevE.79.040101} {\bibfield
  {journal} {\bibinfo  {journal} {Phys. Rev. E}\ }\textbf {\bibinfo {volume}
  {79}},\ \bibinfo {pages} {040101} (\bibinfo {year} {2009})}\BibitemShut
  {NoStop}%
\bibitem [{\citenamefont {Le~Doussal}\ and\ \citenamefont
  {Radzihovsky}(1992)}]{LeDoussal:1992xh}%
  \BibitemOpen
  \bibfield  {author} {\bibinfo {author} {\bibfnamefont {P.}~\bibnamefont
  {Le~Doussal}}\ and\ \bibinfo {author} {\bibfnamefont {L.}~\bibnamefont
  {Radzihovsky}},\ }\href {\doibase 10.1103/PhysRevLett.69.1209} {\bibfield
  {journal} {\bibinfo  {journal} {Phys. Rev. Lett.}\ }\textbf {\bibinfo
  {volume} {69}},\ \bibinfo {pages} {1209} (\bibinfo {year} {1992})},\ \Eprint
  {http://arxiv.org/abs/cond-mat/9208023} {arXiv:cond-mat/9208023} \BibitemShut
  {NoStop}%
\bibitem [{\citenamefont {Gazit}(2009)}]{PhysRevE.80.041117}%
  \BibitemOpen
  \bibfield  {author} {\bibinfo {author} {\bibfnamefont {D.}~\bibnamefont
  {Gazit}},\ }\href {\doibase 10.1103/PhysRevE.80.041117} {\bibfield  {journal}
  {\bibinfo  {journal} {Phys. Rev. E}\ }\textbf {\bibinfo {volume} {80}},\
  \bibinfo {pages} {041117} (\bibinfo {year} {2009})}\BibitemShut {NoStop}%
\bibitem [{\citenamefont {{Le Doussal}}\ and\ \citenamefont
  {Radzihovsky}(2018)}]{LEDOUSSAL2018340}%
  \BibitemOpen
  \bibfield  {author} {\bibinfo {author} {\bibfnamefont {P.}~\bibnamefont {{Le
  Doussal}}}\ and\ \bibinfo {author} {\bibfnamefont {L.}~\bibnamefont
  {Radzihovsky}},\ }\href {\doibase https://doi.org/10.1016/j.aop.2017.08.033}
  {\bibfield  {journal} {\bibinfo  {journal} {Annals of Physics}\ }\textbf
  {\bibinfo {volume} {392}},\ \bibinfo {pages} {340} (\bibinfo {year}
  {2018})},\ \Eprint {http://arxiv.org/abs/1708.05723} {arXiv:1708.05723
  [cond-mat.soft]} \BibitemShut {NoStop}%
\bibitem [{\citenamefont {Zakharchenko}\ \emph {et~al.}(2010)\citenamefont
  {Zakharchenko}, \citenamefont {Rold\'an}, \citenamefont {Fasolino},\ and\
  \citenamefont {Katsnelson}}]{PhysRevB.82.125435}%
  \BibitemOpen
  \bibfield  {author} {\bibinfo {author} {\bibfnamefont {K.~V.}\ \bibnamefont
  {Zakharchenko}}, \bibinfo {author} {\bibfnamefont {R.}~\bibnamefont
  {Rold\'an}}, \bibinfo {author} {\bibfnamefont {A.}~\bibnamefont {Fasolino}},
  \ and\ \bibinfo {author} {\bibfnamefont {M.~I.}\ \bibnamefont {Katsnelson}},\
  }\href {\doibase 10.1103/PhysRevB.82.125435} {\bibfield  {journal} {\bibinfo
  {journal} {Phys. Rev. B}\ }\textbf {\bibinfo {volume} {82}},\ \bibinfo
  {pages} {125435} (\bibinfo {year} {2010})}\BibitemShut {NoStop}%
\bibitem [{\citenamefont {Rold\'an}\ \emph {et~al.}(2011)\citenamefont
  {Rold\'an}, \citenamefont {Fasolino}, \citenamefont {Zakharchenko},\ and\
  \citenamefont {Katsnelson}}]{PhysRevB.83.174104}%
  \BibitemOpen
  \bibfield  {author} {\bibinfo {author} {\bibfnamefont {R.}~\bibnamefont
  {Rold\'an}}, \bibinfo {author} {\bibfnamefont {A.}~\bibnamefont {Fasolino}},
  \bibinfo {author} {\bibfnamefont {K.~V.}\ \bibnamefont {Zakharchenko}}, \
  and\ \bibinfo {author} {\bibfnamefont {M.~I.}\ \bibnamefont {Katsnelson}},\
  }\href {\doibase 10.1103/PhysRevB.83.174104} {\bibfield  {journal} {\bibinfo
  {journal} {Phys. Rev. B}\ }\textbf {\bibinfo {volume} {83}},\ \bibinfo
  {pages} {174104} (\bibinfo {year} {2011})}\BibitemShut {NoStop}%
\bibitem [{\citenamefont {Zhang}\ \emph {et~al.}(1993)\citenamefont {Zhang},
  \citenamefont {Davis},\ and\ \citenamefont {Kroll}}]{PhysRevE.48.R651}%
  \BibitemOpen
  \bibfield  {author} {\bibinfo {author} {\bibfnamefont {Z.}~\bibnamefont
  {Zhang}}, \bibinfo {author} {\bibfnamefont {H.~T.}\ \bibnamefont {Davis}}, \
  and\ \bibinfo {author} {\bibfnamefont {D.~M.}\ \bibnamefont {Kroll}},\ }\href
  {\doibase 10.1103/PhysRevE.48.R651} {\bibfield  {journal} {\bibinfo
  {journal} {Phys. Rev. E}\ }\textbf {\bibinfo {volume} {48}},\ \bibinfo
  {pages} {R651} (\bibinfo {year} {1993})}\BibitemShut {NoStop}%
\bibitem [{\citenamefont {Bowick}\ \emph {et~al.}(1996)\citenamefont {Bowick},
  \citenamefont {Catterall}, \citenamefont {Falcioni}, \citenamefont
  {Thorleifsson},\ and\ \citenamefont {Anagnostopoulos}}]{Bowick:1996wz}%
  \BibitemOpen
  \bibfield  {author} {\bibinfo {author} {\bibfnamefont {M.~J.}\ \bibnamefont
  {Bowick}}, \bibinfo {author} {\bibfnamefont {S.~M.}\ \bibnamefont
  {Catterall}}, \bibinfo {author} {\bibfnamefont {M.}~\bibnamefont {Falcioni}},
  \bibinfo {author} {\bibfnamefont {G.}~\bibnamefont {Thorleifsson}}, \ and\
  \bibinfo {author} {\bibfnamefont {K.~N.}\ \bibnamefont {Anagnostopoulos}},\
  }\href {\doibase 10.1051/jp1:1996139} {\bibfield  {journal} {\bibinfo
  {journal} {J. Phys. I(France)}\ }\textbf {\bibinfo {volume} {6}},\ \bibinfo
  {pages} {1321} (\bibinfo {year} {1996})},\ \Eprint
  {http://arxiv.org/abs/cond-mat/9603157} {arXiv:cond-mat/9603157} \BibitemShut
  {NoStop}%
\bibitem [{\citenamefont {Tr\"oster}(2013)}]{PhysRevB.87.104112}%
  \BibitemOpen
  \bibfield  {author} {\bibinfo {author} {\bibfnamefont {A.}~\bibnamefont
  {Tr\"oster}},\ }\href {\doibase 10.1103/PhysRevB.87.104112} {\bibfield
  {journal} {\bibinfo  {journal} {Phys. Rev. B}\ }\textbf {\bibinfo {volume}
  {87}},\ \bibinfo {pages} {104112} (\bibinfo {year} {2013})}\BibitemShut
  {NoStop}%
\bibitem [{\citenamefont {{Nelson, D.R.}}\ and\ \citenamefont {{Peliti,
  L.}}(1987)}]{NelsonPeliti}%
  \BibitemOpen
  \bibfield  {author} {\bibinfo {author} {\bibnamefont {{Nelson, D.R.}}}\ and\
  \bibinfo {author} {\bibnamefont {{Peliti, L.}}},\ }\href {\doibase
  10.1051/jphys:019870048070108500} {\bibfield  {journal} {\bibinfo  {journal}
  {J. Phys. France}\ }\textbf {\bibinfo {volume} {48}},\ \bibinfo {pages}
  {1085} (\bibinfo {year} {1987})}\BibitemShut {NoStop}%
\bibitem [{\citenamefont {Aronovitz}\ and\ \citenamefont
  {Lubensky}(1988)}]{Aronovitz:1988zz}%
  \BibitemOpen
  \bibfield  {author} {\bibinfo {author} {\bibfnamefont {J.~A.}\ \bibnamefont
  {Aronovitz}}\ and\ \bibinfo {author} {\bibfnamefont {T.~C.}\ \bibnamefont
  {Lubensky}},\ }\href {\doibase 10.1103/PhysRevLett.60.2634} {\bibfield
  {journal} {\bibinfo  {journal} {Phys. Rev. Lett.}\ }\textbf {\bibinfo
  {volume} {60}},\ \bibinfo {pages} {2634} (\bibinfo {year}
  {1988})}\BibitemShut {NoStop}%
\bibitem [{\citenamefont {Coquand}\ \emph {et~al.}(2020)\citenamefont
  {Coquand}, \citenamefont {Mouhanna},\ and\ \citenamefont
  {Teber}}]{Coquand:2020tgb}%
  \BibitemOpen
  \bibfield  {author} {\bibinfo {author} {\bibfnamefont {O.}~\bibnamefont
  {Coquand}}, \bibinfo {author} {\bibfnamefont {D.}~\bibnamefont {Mouhanna}}, \
  and\ \bibinfo {author} {\bibfnamefont {S.}~\bibnamefont {Teber}},\ }\href
  {\doibase 10.1103/PhysRevE.101.062104} {\bibfield  {journal} {\bibinfo
  {journal} {Phys. Rev. E}\ }\textbf {\bibinfo {volume} {101}},\ \bibinfo
  {pages} {062104} (\bibinfo {year} {2020})},\ \Eprint
  {http://arxiv.org/abs/2003.13973} {arXiv:2003.13973 [cond-mat.stat-mech]}
  \BibitemShut {NoStop}%
\bibitem [{\citenamefont {Mauri}\ and\ \citenamefont
  {Katsnelson}(2021)}]{Mauri:2021ili}%
  \BibitemOpen
  \bibfield  {author} {\bibinfo {author} {\bibfnamefont {A.}~\bibnamefont
  {Mauri}}\ and\ \bibinfo {author} {\bibfnamefont {M.~I.}\ \bibnamefont
  {Katsnelson}},\ }\href {\doibase 10.1016/j.nuclphysb.2021.115482} {\bibfield
  {journal} {\bibinfo  {journal} {Nucl. Phys. B}\ }\textbf {\bibinfo {volume}
  {969}},\ \bibinfo {pages} {115482} (\bibinfo {year} {2021})},\ \Eprint
  {http://arxiv.org/abs/2104.06859} {arXiv:2104.06859 [cond-mat.stat-mech]}
  \BibitemShut {NoStop}%
\bibitem [{\citenamefont {Metayer}\ \emph {et~al.}(2021)\citenamefont
  {Metayer}, \citenamefont {Mouhanna},\ and\ \citenamefont
  {Teber}}]{Metayer:2021kxm}%
  \BibitemOpen
  \bibfield  {author} {\bibinfo {author} {\bibfnamefont {S.}~\bibnamefont
  {Metayer}}, \bibinfo {author} {\bibfnamefont {D.}~\bibnamefont {Mouhanna}}, \
  and\ \bibinfo {author} {\bibfnamefont {S.}~\bibnamefont {Teber}},\
  }\href@noop {} {\  (\bibinfo {year} {2021})},\ \Eprint
  {http://arxiv.org/abs/2109.03796} {arXiv:2109.03796 [cond-mat.stat-mech]}
  \BibitemShut {NoStop}%
\bibitem [{\citenamefont {{Guitter, E.}}\ \emph {et~al.}(1989)\citenamefont
  {{Guitter, E.}}, \citenamefont {{David, F.}}, \citenamefont {{Leibler, S.}},\
  and\ \citenamefont {{Peliti, L.}}}]{guitter:1989tbpm}%
  \BibitemOpen
  \bibfield  {author} {\bibinfo {author} {\bibnamefont {{Guitter, E.}}},
  \bibinfo {author} {\bibnamefont {{David, F.}}}, \bibinfo {author}
  {\bibnamefont {{Leibler, S.}}}, \ and\ \bibinfo {author} {\bibnamefont
  {{Peliti, L.}}},\ }\href {\doibase 10.1051/jphys:0198900500140178700}
  {\bibfield  {journal} {\bibinfo  {journal} {J. Phys. France}\ }\textbf
  {\bibinfo {volume} {50}},\ \bibinfo {pages} {1787} (\bibinfo {year}
  {1989})}\BibitemShut {NoStop}%
\bibitem [{\citenamefont {Tentyukov}\ and\ \citenamefont
  {Fleischer}(2000)}]{Tentyukov:1999is}%
  \BibitemOpen
  \bibfield  {author} {\bibinfo {author} {\bibfnamefont {M.}~\bibnamefont
  {Tentyukov}}\ and\ \bibinfo {author} {\bibfnamefont {J.}~\bibnamefont
  {Fleischer}},\ }\href {\doibase 10.1016/S0010-4655(00)00147-8} {\bibfield
  {journal} {\bibinfo  {journal} {Comput. Phys. Commun.}\ }\textbf {\bibinfo
  {volume} {132}},\ \bibinfo {pages} {124} (\bibinfo {year} {2000})},\ \Eprint
  {http://arxiv.org/abs/hep-ph/9904258} {arXiv:hep-ph/9904258} \BibitemShut
  {NoStop}%
\bibitem [{\citenamefont {Ruijl}\ \emph {et~al.}(2020)\citenamefont {Ruijl},
  \citenamefont {Ueda},\ and\ \citenamefont {Vermaseren}}]{Ruijl:2017cxj}%
  \BibitemOpen
  \bibfield  {author} {\bibinfo {author} {\bibfnamefont {B.}~\bibnamefont
  {Ruijl}}, \bibinfo {author} {\bibfnamefont {T.}~\bibnamefont {Ueda}}, \ and\
  \bibinfo {author} {\bibfnamefont {J.~A.~M.}\ \bibnamefont {Vermaseren}},\
  }\href {\doibase 10.1016/j.cpc.2020.107198} {\bibfield  {journal} {\bibinfo
  {journal} {Comput. Phys. Commun.}\ }\textbf {\bibinfo {volume} {253}},\
  \bibinfo {pages} {107198} (\bibinfo {year} {2020})},\ \Eprint
  {http://arxiv.org/abs/1704.06650} {arXiv:1704.06650 [hep-ph]} \BibitemShut
  {NoStop}%
\bibitem [{\citenamefont {Larin}\ and\ \citenamefont
  {Vermaseren}(1993)}]{Larin:1993tp}%
  \BibitemOpen
  \bibfield  {author} {\bibinfo {author} {\bibfnamefont {S.~A.}\ \bibnamefont
  {Larin}}\ and\ \bibinfo {author} {\bibfnamefont {J.~A.~M.}\ \bibnamefont
  {Vermaseren}},\ }\href {\doibase 10.1016/0370-2693(93)91441-O} {\bibfield
  {journal} {\bibinfo  {journal} {Phys. Lett. B}\ }\textbf {\bibinfo {volume}
  {303}},\ \bibinfo {pages} {334} (\bibinfo {year} {1993})},\ \Eprint
  {http://arxiv.org/abs/hep-ph/9302208} {arXiv:hep-ph/9302208} \BibitemShut
  {NoStop}%
\bibitem [{\citenamefont {Bednyakov}\ \emph {et~al.}(2013)\citenamefont
  {Bednyakov}, \citenamefont {Pikelner},\ and\ \citenamefont
  {Velizhanin}}]{Bednyakov:2012rb}%
  \BibitemOpen
  \bibfield  {author} {\bibinfo {author} {\bibfnamefont {A.~V.}\ \bibnamefont
  {Bednyakov}}, \bibinfo {author} {\bibfnamefont {A.~F.}\ \bibnamefont
  {Pikelner}}, \ and\ \bibinfo {author} {\bibfnamefont {V.~N.}\ \bibnamefont
  {Velizhanin}},\ }\href {\doibase 10.1007/JHEP01(2013)017} {\bibfield
  {journal} {\bibinfo  {journal} {JHEP}\ }\textbf {\bibinfo {volume} {01}},\
  \bibinfo {pages} {017} (\bibinfo {year} {2013})},\ \Eprint
  {http://arxiv.org/abs/1210.6873} {arXiv:1210.6873 [hep-ph]} \BibitemShut
  {NoStop}%
\end{thebibliography}%
\appendix
\section{Analytical four-loop results for critical exponents}
\label{sec:an-res-eta}
\begin{widetext}
\begin{align}
  \eta_3 & = \frac{20 \ep}{\denA} + \left(\frac{2800}{\denA^3} + \frac{1060}{3 \denA^2}
           - \frac{74}{3 \denA}\right) \ep^2 \nonumber\\
         & + 
           \left(
           \frac{784000}{\denA^5}
           - \frac{40 (615553 - 591624 \zeta_3)}{27 \denA^4}
           + \frac{2 (1024193 - 1006344 \zeta_3)}{27 \denA^3}
           - \frac{2 (17105 - 20736 \zeta_3)}{27 \denA^2}
           - \frac{155}{9 \denA}
           \right) \ep^3 \nonumber\\
         & + \left(
           \frac{274400000}{\denA^7}
           - \frac{28000 (648943 - 591624 \zeta_3)}{27 \denA^6}
           - \frac{40 (63897618439 + 174575927736 \zeta_3 - 263951628480 \zeta_5)}{
           243 \denA^5} \right.\nonumber\\
         & + \frac{4 (226859519881 + 611469803304 \zeta_3 + 239607720 \zeta_4 - 
           927893517120 \zeta_5)}{729 \denA^4} \nonumber\\
         & - \frac{2 (15308397193 + 40857079644 \zeta_3 + 40756932 \zeta_4 - 62189551440 \zeta_5)}{
           729 \denA^3} \nonumber\\
         & \left.+ \frac{24880019 + 65141136 \zeta_3 + 186624 \zeta_4 - 99921600 \zeta_5}{81 \denA^2}
           - \frac{769 - 2160 \zeta_3}{54 \denA}
           \right) \ep^4 + \mathcal{O}(\ep^5)   \label{eq:eta3-4loop}
\end{align}
\begin{align}
  \eta_4 & =
           \frac{24 \ep}{\denB}
           + \left( \frac{2880}{\denB^3} + \frac{456}{\denB^2} - \frac{24}{24 + d_c}\right) \ep^2 \nonumber\\
         & + \left(
           \frac{691200}{\denB^5}
           - \frac{576 (234137 - 192096 \zeta_3)}{125 \denB^4}
           + \frac{8 (1031777 - 923616 \zeta_3)}{125 \denB^3} 
           - \frac{4 (39029 - 86832 \zeta_3)}{375 \denB^2}
           - \frac{64}{3 \denB}
           \right) \ep^3 \nonumber\\  
         & +  \left(
           \frac{207360000}{\denB^7}
           - \frac{165888 (20501 - 16008 \zeta_3)}{5 \denB^6}
           - \frac{32 (1174399340197 + 3188610294336 \zeta_3 - 4827670269120 \zeta_5)}{1875 \denB^5}
           \right.\nonumber \\
         & + \frac{4 (2761899037843 + 7430870367648 \zeta_3 + 1867173120 \zeta_4 - 11277698973120 \zeta_5)}{5625 \denB^4}
           \nonumber \\
         & - \frac{2 (52639017319 + 140359656168 \zeta_3 + 83125440 \zeta_4 - 213590260800 \zeta_5)}{1875 \denB^3}
           \nonumber \\
         & \left.+ \frac{2 (1074978101 + 2807145072 \zeta_3 + 3907440 \zeta_4 - 4302849600 \zeta_5)}{5625 \denB^2}
           - \frac{16 (13 - 27 \zeta_3)}{9 \denB}
           \right) \ep^4
           + \mathcal{O}(\ep^5) \label{eq:eta4-4loop}
\end{align}
\end{widetext}
\end{document}